\documentclass[12pt]{article}
\usepackage{latexsym}
\usepackage{amsfonts}
\usepackage{epsfig}

\begin{document}
\def\op#1{\mathcal{#1}}
\def\eq#1{(\ref{#1})}

\def\IC{\relax\,\hbox{$\inbar\kern-.3em{\rm C}$}}
\def\inbar{\vrule height1.5ex width.4pt depth0pt}

\def\a{\alpha}
\def\b{\beta}
\def\bfone{\relax{\rm 1\kern-.35em 1}}
\def\bfnull{\relax{\rm O \kern-.635em 0}}
\def\bos{{\rm bos}}
\def\c{\chi}
\def\cb{\bar{\chi}}
\def\cc{{\rm c.c.}}
\def\d{\delta}
\def\D{\Delta}
\def\der{\partial}
\def\dop{{\rm d}\hskip -1pt}
\def\dx{\right}
\def\e{\epsilon}
\def\g{\gamma}
\def\G{\Gamma}
\def\i{\imath}
\def\im{{\rm Im}\op{N}}
\def\imez{\frac{{\rm i}}{2}}
\def\l{\lambda}
\def\L{\Lambda}
\def\m{\mu}
\def\mez{\frac{1}{2}}
\def\n{\nu}
\def\na{\nabla}
\def\o{\omega}
\def\O{\Omega}
\def\ol{\overline}
\def\p{\psi}
\def\qu{\frac{1}{4}}
\def\r{\rho}
\def\re{{\rm Re}\op{N}}
\def\s{\sigma}
\def\S{\Sigma}
\def\sx{\left}
\def\t{\tau}
\def\th{\theta}
\def\Th{\Theta}
\def\ve{\varepsilon}
\def\z{\zeta}

\newcommand{\be}{\begin{equation}}
\newcommand{\ee}{\end{equation}}
\newcommand{\ba}{\begin{eqnarray}}
\newcommand{\ea}{\end{eqnarray}}
\newcommand{\ban}{\begin{eqnarray*}}
\newcommand{\ean}{\end{eqnarray*}}
\newcommand{\nn}{\nonumber}
\newcommand{\noi}{\noindent}
\newcommand{\fgl}{\mathfrak{gl}}
\newcommand{\fu}{\mathfrak{u}}
\newcommand{\fsl}{\mathfrak{sl}}
\newcommand{\fsp}{\mathfrak{sp}}
\newcommand{\fusp}{\mathfrak{usp}}
\newcommand{\fsu}{\mathfrak{su}}
\newcommand{\fp}{\mathfrak{p}}
\newcommand{\fso}{\mathfrak{so}}
\newcommand{\fg}{\mathfrak{g}}
\newcommand{\fr}{\mathfrak{r}}
\newcommand{\fe}{\mathfrak{e}}
\newcommand{\rE}{\mathrm{E}}
\newcommand{\rSp}{\mathrm{Sp}}
\newcommand{\rSO}{\mathrm{SO}}
\newcommand{\rSL}{\mathrm{SL}}
\newcommand{\rSU}{\mathrm{SU}}
\newcommand{\rUSp}{\mathrm{USp}}
\newcommand{\rU}{\mathrm{U}}
\newcommand{\rF}{\mathrm{F}}
\newcommand{\R}{\mathbb{R}}
\newcommand{\C}{\mathbb{C}}
\newcommand{\Z}{\mathbb{Z}}
\newcommand{\Hb}{\mathbb{H}}
\def\N\mathcal{N}


\begin{titlepage}
\rightline{CERN-PH-TH/2004-194} \vskip 1cm

\begin{center}
{\LARGE { Dyonic Masses from Conformal Field Strengths in D even Dimensions.}}\\
\vskip 1.5cm
  {\bf Riccardo D'Auria$^{a,1}$ and Sergio Ferrara$^{b,2}$} \\
\vskip 0.5cm
\end{center}
\begin{center}
{\small $^{a}$ Dipartimento di Fisica, Politecnico di
Torino,\\ Corso Duca degli Abruzzi 24, I-10129 Torino, Italy}\\
 and\\
{\small  Istituto Nazionale di Fisica Nucleare (INFN) - Sezione di
Torino,\\ Via P. Giuria 1, I-10125 Torino, Italy}\\
\vskip 0.5cm
 {\small {$^{b}$ CERN, Theoretical Physics Department, Geneva,
 Switzerland}}
\end{center}
\begin{center}
e-mail: {$^{1}$riccardo.dauria@polito.it\quad $^{2}$
sergio.ferrara@cern.ch}
\end{center}
\vskip 1cm
\begin{abstract}

We show that D/2--form gauge fields in D even dimensions can get a
mass with both electric and magnetic contributions when coupled to
conformal field--strengths whose gauge potentials are $\frac
{D-2}{2}$-- forms. Denoting by $e^I_\L$ and $m^{I\L}$ the electric
and magnetic couplings, gauge invariance requires: $ e^I_\L
m^{J\L}\mp e^J_\L m^{I\L}=0$, where $I,\L= 1\cdots m$ denote the
species of gauge potentials of degree $D/2$ and gauge fields of
degree $D/2-1$, respectively. The minus and plus signs refer to
the two different cases $D=4n$ and $D=4n+2$ respectively and the
given constraints are  respectively ${\rm {Sp}}(2m)$ and ${\rm
{O}}(m,m)$ invariant. For the simplest examples, ($I,\L=1$ for
$D=4n$ and $I,\L=1,2$ for $D=4n+2$) both the $e,m$ quantum numbers
contribute to the mass  $\m=\sqrt {e^2 +m^2}$. This phenomenon
generalizes to $D$ even dimensions the coupling of massive
antisymmetric tensors which appear in $D=4$ supergravity
Lagrangians, which derive from flux compactifications in higher
dimensions. For $D=4$ we give the supersymmetric generalization of
such couplings using $N=1$ superspace.
\end{abstract}

\end{titlepage}

\section{Introduction}
\label{sec:intro} It is well known that in $D=4$ dimensions a
massless antisymmetric tensor $B_{\m\n}$ is "dual" to a massless
scalar, while in the massive case \cite{Takahashi:1970ev} it is
dual to a massive vector. In this note we consider a
generalization of this phenomenon in $D$ even dimensions when a
$D/2$--form receives contribution to the mass both from an
"electric" and a "magnetic" type of coupling. Denoting by $e^I_\L$
and $m^{I \L}$ the "electric" charges and the "magnetic" charges
respectively, such couplings are of the type $e^I_\L F^\L \wedge
B_I$, where $F^\L=dA^\L (I,\L=1\cdots m)$ are conformal
field--strengths \cite{Ferrara:1998bv}  and $\frac {1}{2}m^{I \L}
{^*}F_\L \wedge B_I$ \cite{Romans:1985tw}. Here and in the
following we use the differential form language.

These two couplings have in fact different origin, the former
being the four dimensional version of the Green--Schwarz
(geometrical) coupling \cite{Dine:1987xk} being gauge invariant
under $\d A^\L=d \phi^\L$ and $\d B_I = d\L_I$. The latter comes
from certain flux or Scherk--Schwarz compactifications, and it can
be made gauge invariant (when $e^I_\L=0$) only if  $\d
A^\L=-m^{I\L}\L_I$ and a $B_{I\,\m\n}$ mass term $\frac {1}{2}
m^{I\L} m^{J\L} B_I\wedge ^*B_J$
is added \cite{Louis:2002ny}.\\
However it has recently been noted in compactification of Type IIB
with one tensor field
 \cite{Louis:2002ny} and, more generally,  in the context of $N=2$,
$D=4$ supergravity coupled to vector multiplets and (an arbitrary
number of) tensor multiplets \cite{D'Auria:2004yi}, that one can
have both "electric" and "magnetic" type of mass terms, namely the
linearized Lagrangian has the following form\footnote{We used the
fact that for a $k$--form $\o^{(k)}$ we have $\o^{(k)} \wedge {}^*
\o^{(k)}=(-1)^{D-t} \, k! \, \o_{\m_1 \dots \m_k} \o^{\m_1 \dots
\m_k} \sqrt{|g|} \, dx^1 \wedge\dots\wedge dx^D $, where $t$ is
the space-time signature of the metric chosen to be "mostly
minus".}
 :\begin{eqnarray}\label{Lag}
    \mathcal {L}^{(D=4)} &=& -\frac {1}{2}H_I \wedge ^{* }H_I + \frac {1}{2} m^{\L I} m^{\L J}\left(
    B_I+ (m^{-1})_{\G I} F^{\G}\right)\wedge ^*\left(
    B_J+ (m^{-1})_{\D J} F^\D\right) \nonumber\\ &&- \frac {1}{2} e^I_\L m^{\L J}B_I
    \wedge B_J - e^I_\L B_I \wedge F^\L.
\end{eqnarray}
It is immediate to verify that this Lagrangian is invariant under
the gauge transformations \begin{equation}\label{tras}\d
A^\L=-m^{I\L}\L_I, \quad, \quad \d B_I = d\L_I
\end{equation} provided
the condition
\begin{equation}\label{constr1} e^I_\L m^{J\L}- e^J_\L m^{I\L}=0
\end{equation}
is satisfied \cite{Louis:2002ny,D'Auria:2004yi} (this condition is
void in reference \cite{Louis:2002ny} since in that case $I=1$).
Note that this condition is ${\rm {Sp}}(2m)$ invariant.

\noi In the following we discuss the generalization of this $D=4$
case to any even $D$ space--time dimensions.

We first observe that for $D=4n$, $n>1$, the above properties of
the theory remain the same as in the $n=1$ case. This appears to
be evident if one uses the differential forms language. On the
other hand, for $D=4n+2$, the combined presence of electric and
magnetic mass terms, requires now the condition:
 \begin{equation}\label{constr2} e^I_\L
m^{J\L}+ e^J_\L m^{I\L}=0.
\end{equation}
This is so because the term $-\frac {1}{2} e^I_\L m^{J\L}B^I
\wedge B_J$ whose variation must cancel the variation of the term
$-e^I_\L B_I \wedge F^\L$ under the combined gauge
transformation:\begin{equation}\label{comb} \d B_I=d \L_I,
\quad,\quad \d A^\L=-m^{i\L} \L_I
\end{equation}
is symmetric for $D=4n$ and antisymmetric for $D=4n+2$. This
explains why for $D=4n$ we can have $I,\L=1$, but for $D=4n+2$ the
simplest case is $I,\L = 1,2$ with $e^I_\L = e \epsilon^I_\L$,
where $\epsilon =- (\epsilon)^T, \epsilon^2=-1$, and $m^{I\L} =m
\d^{I\L}$. Only when $m^{I\L}=0$  we have no restriction on the
$e^I_\L$ and we can also $I,\L=1$ also for $D=4n+2$. Note that
equation (\ref{constr2} is ${\rm {O}}(m,m)$ invariant. The
difference in the invariance groups in the equations
(\ref{constr1}) and (\ref{constr2})is related to the duality
rotations of conformal field--strengths of degree $D/2$ for rhe
$d=4n$ and $D=4n+2$. Gauge invariant mass terms also exist in odd
dimensions as is discussed in \cite{Townsend:1983xs}.

\section{Massive gauge fields in even dimensions}

Let us extend the Lagrangian (\ref{Lag}) of the four dimensional
case. If we take $D=4n$ the Lagrangian has exactly the same form
where now $F^\L=dA^\L$ and $B_I$ are $2n$--forms and $H_I=d B_I$
are $2n+1$--forms. In the simplest case discussed above, namely if
 we take just one $F$ field--strength and one $B$ gauge field, the
 Lagrangian takes the form:\begin{eqnarray}\label{Lag2}
 \mathcal {L} = -\frac {1}{2}H \wedge ^{* }H + \frac {1}{2} m^2\left(
    B+ (m^{-1}) F\right)\wedge ^*\left(
    B+ (m^{-1}) F\right) \\ \nonumber - \frac {1}{2}e m\left( B+ (m^{-1}) F
    \right)\wedge \left(B+ (m^{-1}) F \right).
\end{eqnarray}
where we have added the total derivative $-\frac {e}{2m}F \wedge
F$. In this form the Lagrangian is manifestly invariant under the
gauge transformations (\ref{tras}) since the quantity $B+ m^{-1}
F$ is gauge invariant. By redefining
\begin{equation}\label{red} B + m^{-1}dA \longrightarrow B
\end{equation}
 the Lagrangian takes the simplest form:
\begin{equation}\label{Lag3}
 {\mathcal {L}}^{D=4n}= -\frac {1}{2}H \wedge ^{* }H + \frac {1}{2} m^2 B \wedge ^*
    B  - \frac {1}{2} e m B \wedge B.
\end{equation}
For $D=4n+2$ we also take the simplest case, namely we consider
two $F$'s and two $B$'s setting $I,\L = 1,2$ with $e^I_\L = e
\epsilon^I_\L$ and $m^{I\L} =m \d^{I\L}$. For notational
simplicity we do not write the indices $I,\L = 1,2$ explicitly,
treating $F$ and  $B$ as two--dimensional vectors and
$\epsilon^I_\L$ as the $2 \times 2$ antisymmetric matrix
$\epsilon, \epsilon^2=-1$. Adding as before the total derivative
$-\frac {e}{2m}F^T \wedge \epsilon F$, the Lagrangian for
$D=4n+2$, after the redefinition (\ref{red}) has been performed,
takes the following form:.
\begin{equation}\label{Lag4}
 -{\mathcal {L}}^{D=4n+2} = -\frac {1}{2}H^T \wedge ^{* }H + \frac {1}{2} m^2 B^T \wedge ^*
    B  - \frac {1}{2} e m B^T \wedge \epsilon B.
\end{equation}
where the overall minus sign with respect to the $D=4n$ case is
due to the requirement of positive kinetic energy and mass
squared.\\
\noi \, From the Lagrangians (\ref{Lag3}),(\ref{Lag4}) we obtain
the following equations of motion respectively:
\begin{equation}\label{eom1}
d ^{*} d B+m^2 {^*} B -e m B =0  \quad, \quad D=4n
\end{equation}
\begin{equation}\label{eom2}
d ^* d B- m^2 {^*} B +e m \epsilon B =0\quad, \quad D=4n+2
\end{equation}
The integrability conditions of these equations can be written as
the transversality conditions
\begin{equation}\label{int1}
d\, ^* \left( B+ \frac {e}{m} {^{*}} B \right)=0\quad, \quad D=4n
\end{equation}
\begin{equation}\label{int2}
d \,^{*} \left( B- \frac {e}{m}  \epsilon \,{^*}B\right)=0\quad,
\quad D=4n+2
\end{equation}
Using the constraints (\ref{int1}), (\ref{int2}) we can write the
equations (\ref{eom1}),(\ref{eom2}) as equations for $^*B$  and,
taking linear combinations, we obtain:
\begin{equation}\label{eom3} d {^*} d
\left(B+\frac {e}{m} {^*} B \right) +(e^2 + m^2) ^* \left(B+\frac
{e}{m} {^*} B \right)  =0  \quad, \quad D=4n
\end{equation}
\begin{equation}\label{eom4} d {^*} d
\left(B-\frac {e}{m} {^*} \epsilon B \right) +(e^2 + m^2) ^*
\left(B-\frac {e}{m} {^*} \epsilon B \right)  =0  \quad, \quad
D=4n+2
\end{equation}
In deriving these equations we have used the property that, in
even dimensions,  the Hodge star operator on a $p$--form
$\o^{(p)}$:
\begin{equation}\label{hodge}
^* {\omega}^{(p)} =\frac {1} {(D-p)!}\o _{\m_1\cdots \m_p}\sqrt
{-g}{\epsilon^{\m_1\cdots \m_p}_{\quad \quad \n_1\cdots
\n_{D-p}}}d x^\n_1\wedge \cdots \wedge d x^{\n_{D-p}}
\end{equation}
satisfies the relation
\begin{equation}\label{Hodge} ^{**} \o^{(p)}= (-1)^{p+1} \o^{(p)}
\end{equation}
Taking the Hodge-- star of equations (\ref{eom3}), (\ref{eom4})
and recalling that the Klein--Gordon operator $\square$ is defined
as
\begin{equation}\label{kg}
\square = \d d+d \d \quad,\quad \d =-{^*}d {^*}
\end{equation}
we obtain the equations of motion in their standard form:
\begin{equation}\label{eom5}
\square \left(B+\frac {e}{m} {^*} B \right) +(e^2 + m^2) ^*
\left(B+\frac {e}{m} {^*} B \right)  =0  \quad, \quad D=4n
\end{equation}
\begin{equation}\label{eom6}
\square \left(B-\frac {e}{m} {^*} \epsilon B \right) +(e^2 + m^2)
^* \left(B-\frac {e}{m} {^*} \epsilon B \right)  =0  \quad, \quad
D=4n+2
\end{equation}
The equations (\ref{eom5}),(\ref{eom6}), together with the
transversality conditions (\ref{int1}),(\ref{int2}), describe a
massive $D/2$--form of mass $\m=\sqrt {e^2+m^2}$. Note that for
$D=4n+2$ the field $B$ is a $2$--vector $B=(B_1,B_2)^T$, each
component being a $(2n+1)$--form.

\section{The dual formulation}

In the massless case it is known, by Poincar\`e duality, that a
massless $D/2$--form is "dual" to a massless $D/2-2$--form. For
example in $D=4$ a $2$--form is dual to a scalar and in $D=6$ a
$3$--form is dual to a vector.

Here we show that a massive $D/2$--form, described in the previous
section, is dual to a massive $D/2-1$--form. As previously
discussed a doubling of the $D/2$ gauge potential is required for
$D=4n+2$ when both electric and magnetic masses are present in the
theory.

The process of dualization at the level of the Lagrangian requires
the first--order formalism for the gauge potential $B$ and it is a
straightforward generalization of the method used in references
\cite{Cecotti:1987qr,Louis:2002ny}. Let us discuss separately the
two cases $D=4n$ and $D=4n+2$.

\subsection{$D=4n$}
The Lagrangian (\ref{Lag2}) can be dualized by rewriting it in
first order form with $B$ and $H$ independent $(2n)$-- and
$(2n+1)$--forms, and $A$ and $F$ independent $(2n-1)$-- and
$(2n)$--forms, respectively \cite {Cecotti:1987qr}, and enforcing
the relations $H=d B$ and $F=dA$ by suitable Lagrangian
multipliers $\r$ and $\xi$ which are $(2n)$-- and $(2n-1)$--forms
respectively. We start with:
\begin{eqnarray}\label{firstorder1}
    {\mathcal {L}}^{(D=4n)}&=& -\frac {1}{2}H \wedge ^{* }H + \frac {1}{2} m^2 (B+\frac{1}{m} F) \wedge ^*
    (B+\frac{1}{m} F)  \nn\\
    &&- \frac {1}{2} e m (B+\frac{1}{m} F) \wedge (B + \frac{1}{m} F)+\r \wedge \left(H-d(B+\frac{1}{m} F)
    \right)\nn\\
    &&+\xi \wedge (F-dA)
\end{eqnarray}
The original Lagrangian (\ref{Lag2}) is retrieved by imposing the
equations of motion of $\r$ and $\xi$. The dual Lagrangian
${\mathcal {L}}^{D=4n}_{Dual}$ is instead obtained by varying the
forms $H$ and $B$. One obtains:
\begin{eqnarray}\label{dih}
    \frac {\d \mathcal {L}}{\d H}&=&0 \longrightarrow {{}^*}H =-\r
    \rightarrow H=-{{}^*}\r \\
\label{dib} \frac {\d \mathcal {L}}{\d B}&=&0 \longrightarrow m^2
{{}^*} (B+\frac{1}{m} F) - e m (B +\frac{1}{m} F) =d\r\\
\frac {\d \mathcal {L}}{\d F}&=& \xi + m {}^*(B +\frac{1}{m} F) -
e (B+\frac{1}{m} F) =\frac{1}{m} d\r\\
    \frac {\d \mathcal {L}}{\d A}&=& 0 \longrightarrow d\xi = 0
\end{eqnarray}
From the previous equations we easily find
\begin{eqnarray}
\label{xi} \xi &=& 0 \\
B + \frac{1}{m} F &=& - \frac{1}{e^2 + m^2} \left(
\frac{e}{m} d \r + ^*d \r \right)\\
\label{bbstar} {}^*(B + \frac{1}{m} F) &=& - \frac{1}{e^2 + m^2}
\left( \frac{e}{m} {}^*d \r - d \r \right)
\end{eqnarray}
We redefine $(B+\frac{1}{m}F) \longrightarrow B$. Then
Hodge--starring (\ref{dib}) and combining with (\ref{dib}), we
obtain:
\begin{eqnarray}\label{bbst}
B  &=& - \frac {1} {e^2+m^2} \left( \frac {e} {m} d \r + ^*d \r
  \right)\\
^*B  &=& - \frac {1} {e^2+m^2} \left( \frac {e} {m} ^*d \r -d \r
  \right)
\end{eqnarray}
Substituting in (\ref{firstorder1}) one finds:
\begin{equation}\label{ldu1}
\mathcal {L}^{D=4n}_{(Dual)}=\frac {1}{2}\left( \frac
{1}{e^2+m^2}\,d \r \wedge ^* d \r + ^* \r \wedge \r\right)
\end{equation}
which is indeed the Lagrangian for a massive $2n$--form
$\overline{\r}= \rho(e^2+m^2)^{-1/2}$. The equations of motion
are:\begin{equation}\label{eomf1} d^*d \r- (e^2+m^2){^*}\r=o
\end{equation}
together with the transversality condition $d ^{*}\r=0$.  Using
the definition of the Klein--Gordon operator (\ref{kg}) we obtain
in the standard formalism for the divergenceless field $\r$:
\begin{equation}\label{kgd}
    \square \r + (e^2+m^2) \r =0
\end{equation}

\subsection{D=4n+2}

The Lagrangian in first order formalism is now:

\begin{eqnarray}\label{firstorder2}
{\mathcal {L}}^{D=4n+2}&=& \frac {1}{2}H^T \wedge ^{* }H - \frac
{1}{2} m^2 (B +\frac{1}{m} F)^T \wedge ^* (B +\frac{1}{m}F)\nn\\
&& +\frac {1}{2} e\, m (B +\frac{1}{m}F)^T \wedge \e ( B +
\frac{1}{m} F) - \r^T \wedge \left(H-d \,(B+\frac{1}{m}F) \right)\nn\\
&& + \xi^T \wedge (F- d A)
\end{eqnarray}
The original Lagrangian (\ref{Lag4})is retrieved as before from
the equation of motion of $\r$.

\noi To obtain the dual Lagrangian we proceed as in the former
case $D=4n$ by varying $H$, $B$, $F$ and $A$. We obtain:

\begin{eqnarray}\label{dih2}
\frac {\d \mathcal {L}}{\d H}&=&0 \longrightarrow {}^*H =\r
\rightarrow H=-{}^*\r \\
\label{dib2} \frac {\d \mathcal {L}}{\d B}&=&0 \longrightarrow m^2
{}^* (B +\frac{1}{m}F) - e\, m \,\epsilon (B +\frac{1}{m}F) =d\r\\
\frac {\d \mathcal {L}}{\d F}&=&0 \longrightarrow  m {}^*(B +
\frac{1}{m} F) - e\,\epsilon(B +\frac{1}{m} F) +\xi = \frac{1}{m} d\r \\
\frac {\d \mathcal {L}}{\d A}&=&0 \longrightarrow d\xi=0
\end{eqnarray}
Redefining $B+m^{-1}F \longrightarrow B$ and proceeding as before
we find the expressions of $B$ and $^*B$:
\begin{eqnarray}
\label{xi2}\xi &=& 0\\
\label{bbstar2} B  &=&  \frac {1} {e^2+m^2} \left( \frac {e} {m}
\epsilon d \r + ^*d \r
  \right)\\
^*B  &=& - \frac {1} {e^2+m^2} \left( \frac {e} {m} \epsilon ^*d
\r +d \r
  \right)
\end{eqnarray}
Substituting in equation (\ref{firstorder2})we find:
\begin{equation}\label{ldu2}
\mathcal {L}^{D=4n+2}_{(Dual)}=-\frac {1}{2}\left( \frac
{1}{e^2+m^2}\,d \r^T \wedge ^*d \r -^*\r^T \wedge \r \right)
\end{equation}
whose equations of motion  are:
\begin{equation}\label{eomf2}
 d^*d \r+(e^2+m^2){^*}\r=o
\end{equation}
together with the transversality condition $d ^{*}\r=0$.  In the
usual formalism  the equation (\ref{eomf2}) for the divergeless
field $\r$ becomes:
\begin{equation}\label{kg2}
    \square \r + (e^2+m^2) \r =0
\end{equation}
 as before, except
for the fact that $\r$ is now a two--dimensional vector.

\section{Supersymmetric generalization in $D=4$}

 The Lagrangian (\ref{firstorder1}) can be easily generalized to
 $D=4$ $N=1$ superspace by introducing a vector multiplet and a
 linear multiplet potential \cite{Ferrara:1974ac,Siegel:1979ai}:
 \begin{equation}\label{lin}
   V= \overline {V}\, ; \quad \quad L_{\a} \quad ;(\overline {D}_{\dot{\a}} L_{\a})=0
\end{equation}
through which the gauge invariant field strengths can be
constructed \cite{Wess:1992cp} namely:
\begin{eqnarray}\label{super1}
    W_\a &=& \overline {D}^2 D_{\a} V;\\
   \label {super2}  L &=& i\left(D^{\a} L_\a -
    \overline {D}_{\dot{\a}} \overline {L}^{\dot{\a }}\right)
\end{eqnarray}
which are invariant under:
\begin{equation}\label{inv1}
    \d V= \Sigma + \overline {\Sigma},\quad \quad \overline {D} \Sigma
    =0
\end{equation}
\begin{equation}\label{inv2}
    \d L_\a = \overline {D}^2 D_\a U, \quad \quad U=\overline {U}
\end{equation}
respectively. Note that $W^\a$ and $L$ satisfy the Bianchi
identities:
\begin{equation}\label{Bian}
    D^\a W_\a = \overline {D}_{\dot{\a}} \overline {W}^{\dot{\a}},
    \quad \quad D^2L= \overline {D}^2 L=0
\end{equation}
as a consequence of the identity:
\begin{equation}\label{real}
    D^\a \overline{D}^2 D_\a = \overline{D}_{\dot{\a}} D^2 \overline{D}^{\dot{\a}}
\end{equation}
 We observe that the combination
$L_\a + m^{-1} W_\a$ is invariant under:
\begin{equation}\label{gen}
    \d L_\a = \overline {D}^2 D_{\a} U,\quad \quad \d V= -m U
\end{equation}
which is the supersymmetric generalization of the bosonic gauge
invariance (for $I,\L=1$) of equations (\ref {comb}). The physical
degrees of freedom of $L_\a$ are a scalar $\phi$, a 2--form $B$
and a Weyl spinor $\z$, while the vector multiplet
contains a vector $A$ and a Weil spinor $\l$.\\
The supersymmetric generalization of (\ref{Lag2} will be a $N=1$
massive vector multiplet, containing a massive vector, a massive
scalar and a massive Dirac spinor, all with mass $\m =\sqrt {e^2 +
m^2 }$.\\\noi To derive this result we generalize to superspace
the Lagrangian (\ref {firstorder1}) by introducing two Lagrange
multipliers $\psi_\a, (\overline {D}_{\dot{\a}} \psi_\a)=0$ and
$\O, (\O ={^*}\O)$ so that the action is \footnote { Note that in
the case we have several fields $L_I$ and $W_\a^\L$ we easily find
the the gauge invariance under (\ref {inv1}) and (\ref {inv2})
requires the condition (\ref {constr1}).}:
\begin{eqnarray}\label{superlag}
\mathcal {L}&=& \int d^4 \theta \left[- \frac {1} {2} L^2 + \O
\left(L - i \mathcal {D}^\a \left (L_\a
    + m^{-1} W_\a \right)
    +i \mathcal {\overline {D}}_{\dot\a} \left ( \overline {L}^{\dot{\a}}
    + m^{-1} \overline {W}^{\dot{\a}}\right)\right)\right]
    \nonumber\\
    &+& \left [\int d^2 \theta \frac {1}{2} m(m+ie) \left( L^\a +
    m^{-1}W^\a\right) \left(L_\a + m^{-1} W_\a  \right)+i \psi
    \left(W_\a -\overline {D}^2 D_\a V
    \right)\right.\nonumber\\&&\left.
+\,\mbox{h.c.}\phantom{\int} \right].
\end{eqnarray}
 If we vary with
respect to $\psi^\a$ and $\O$ we get:
\begin{eqnarray}\label{multip}
    W_\a &=& \overline {D} ^2 D_\a V \\
    L &=& i \left ( \mathcal {D}^\a \left ( L_\a + m^{-1} W_\a \right)
    - i \mathcal {\overline {D}}_{\dot\a} \left ( \overline {L}^{\dot{\a}}
    + m^{-1} \overline {W}^{\dot{\a}}\right)\right)\\ \nonumber
    =&i&\left
    (\mathcal {D}^\a L_\a - \mathcal {\overline {D}}_{\dot\a}
\overline {L}^{\dot{\a}}
    \right )
\end{eqnarray}
which gives the Lagrangian:
\begin{equation}\label{finlag}
  \mathcal {L}=  -\frac {1} {2}\int d ^4 \theta  L^2 + \frac {1} {2}\left(\int d^2
    \theta\, \,
    m(m+ie) L^\a L_\a + h.c.\right)
\end{equation}
which is the supersymmetric generalization of the Lagrangian (\ref
{Lag2}).

The dual supersymmetric Lagrangian is obtained instead by varying
(\ref {superlag}) with respect to $L$, $L_\a$ and $W_\a$. One
obtains:
\begin{eqnarray}\label{dua}
    &&L=\O,\quad \quad \psi =0 \\
 &&m(m+ie)(L_\a+ m^{-1}W_\a) +i \overline {D}^2 D_\a \O =0
\end{eqnarray}
By insertion of (\ref{dua}) into (\ref {superlag}) one obtains the
dual Lagrangian:
\begin{equation}\label{dususy}
    \mathcal {L}_{Dual} = \frac {1}{2}\int d^4 \theta \O^2 + \left (\frac
    {1}{2}\int d^2\theta \frac {m-ie}{m(m^2+e^2)}
 (\overline {D}^2 D^\a \O)^2 + h.c. \right )
\end{equation}
The last (chiral) term gives:
\begin{equation}\label{chiral}
\mathcal {L}_{Chiral} = \frac {1}{2}\int d^2\theta \frac
{1}{(m^2+e^2)}\left [\left(W^\a W_a +h.c.\right)-i \frac
{e}{m}\left(W^\a W_a -h.c.\right)\right]
\end{equation}
The first term in  (\ref {chiral}) is the kinetic term of a
massive vector superfield $ (e^2 + m^2)^{-\frac {1}{2}}\,W_\a $
while the last term is a total derivative which corresponds to the
supersymetric generalization of the topological term $\frac
{e}{2m} F \wedge F$, with $\theta$ --parameter $\frac {e}{m}$.

\section{Conclusions} We have shown that $D/2$--forms in $D$ even
dimensions with both electric and magnetic mass terms are dual to
massive $D/2-1$--forms with dyonic mass $\m= \sqrt {e^2+m^2}$.
This phenomenon has a supersymmetric generalization in $D=4$ for
$N=1$ and $N=2$. In the latter case a tensor multiplet provides
the dual version of the Higgs mechanism in which a hypermultiplet
is eaten by a vector multiplet to combine into a long massive
multiplet with mass $\m$. Under suitable assumptions on the
$e^I_\L$ and $m^{I\L}$ matrices, in the multivariable case the
squared mass matrix becomes $\m^2= ee^T+ mm^T$.

We observe that, when extended to interactions, the $N=2$
lagrangian with both $e$ and $m$ present are more general than the
ones with either $e$ or $m$ vanishing. In particular they may give
rise to spontaneous supersymmetry breaking when suitably truncated
to $N=1$, as in the case of Calabi--Yau compactifications of Type
IIB on orientifolds
\cite{Taylor:1999ii,Blumenhagen:2003vr,Giddings:2001yu,Giryavets:2003vd}.
In this case the GVW superpotential $W$ \cite{Gukov:1999ya}, with
$e$ and $m$ different from zero may lead to non trivial vacua
solutions with vanishing vacuum energy and $N=1\rightarrow N=0$
supersymmetry breaking. The resulting potential corresponds to an
electric and magnetic Fayet--Iliopoulos term
\cite{Antoniadis:1995vb} whose $N=2$ supergravity generalization
was introduced in reference \cite {Louis:2002ny} for $I=1$ and in
\cite{D'Auria:2004yi} in the general case.

\section{Acknowledgements}
We thank M. Trigiante for useful discussions. One of us (S.F.)
would like to thank the Department of Physics of Politecnico di
Torino where part of this work was performed.
 Work supported in
part by the European Community's Human Potential Program under
contract HPRN-CT-2000-00131 Quantum Space-Time, in which R. D'A.
is associated to Torino University and S. Ferrara is associated
with INFN Frascati National Laboratories. The work of S. F. has
been supported in part by DOE grant DEFG03-91ER40662,Task C.

\end{document}